\documentclass[aps,prd,a4paper,twocolumn,amsmath,showpacs,superscriptaddress,nofootinbib,preprintnumbers]{revtex4-1}

\usepackage{verbatim}
\usepackage[T1]{fontenc}
\usepackage[utf8]{inputenc}
\usepackage[american]{babel}
\usepackage{epsfig}
\usepackage{graphicx}
\usepackage{booktabs}
\usepackage{multirow}
\usepackage{dcolumn}
\usepackage{amsmath}
\usepackage{mathtools}
\usepackage{amsfonts}
\usepackage{amssymb}
\usepackage{epstopdf}
\usepackage{bm}
\usepackage{siunitx}
\usepackage{braket}
\usepackage{enumitem}
\usepackage{soul}
\usepackage[table]{xcolor}
\usepackage{color}
\usepackage{transparent}
\usepackage{pifont}

\newcommand{\beq}{\begin{equation}}
\newcommand{\eeq}{\end{equation}}
\newcommand{\ben}{\begin{eqnarray}}
\newcommand{\een}{\end{eqnarray}}
\newcommand{\bi}{\begin{itemize}}
\newcommand{\ei}{\end{itemize}}

\newcommand{\neff}{N_{\mathrm{eff}}}
\newcommand{\mnu}{{\Sigma}m_{\nu}}

\newcommand{\remove}[1]{}

\def\refe@jnl#1{{#1}}
\def\aj{\refe@jnl{Astron.~J.}}
\def\araa{\refe@jnl{Annu.~Rev.~Astron.~Astrophys.}}
\def\apj{\refe@jnl{Astrophys.~J.}}
\def\apjl{\refe@jnl{Astrophys.~J.~Lett.}}
\def\aap{\refe@jnl{Astron.~Astrophys.}}
\def\mnras{\refe@jnl{Mon.~Not.~R.~Astron.~Soc.}}
\def\prd{\refe@jnl{Phys.~Rev.~D}}
\def\fcp{\refe@jnl{Fund.~Cos.~Phys.}}
\def\physrep{\refe@jnl{Phys.~Rep.}}
\def\physlett{\refe@jnl{Phys.~Lett.}}

\def\pe{{p_{e^-}}}

\def\pe1{{p_{e_1}}}

\begin{document}

\title{ Reducing the $H_0$ and $\sigma_8$ tensions with Dark Matter-neutrino interactions.}

\author{Eleonora Di Valentino}
\email{eleonora.divalentino@manchester.ac.uk}
\affiliation{Jodrell Bank Center for Astrophysics, School of Physics and Astronomy, University of Manchester, Oxford Road, Manchester, M13 9PL, UK}
\author{C\'eline B\oe hm} 
\email{c.m.boehm@durham.ac.uk}
\affiliation{Institute for Particle Physics Phenomenology, Durham University, South Road, Durham, DH1 3LE, United Kingdom}
\affiliation{LAPTH, U. de Savoie, CNRS,  BP 110, 74941 Annecy-Le-Vieux, France}
\affiliation{Perimeter Institute, 31 Caroline St N., Waterloo Ontario, Canada N2L 2Y5}
\author{Eric Hivon} 
\affiliation{Institut d'Astrophysique de Paris (UMR7095: CNRS \& UPMC- Sorbonne Universities), F-75014, Paris, France}
\author{Fran\c cois R. Bouchet} 
\affiliation{Institut d'Astrophysique de Paris (UMR7095: CNRS \& UPMC- Sorbonne Universities), F-75014, Paris, France}

\preprint{}
\begin{abstract}
The introduction of Dark Matter-neutrino interactions modifies the Cosmic Microwave Background (CMB) angular power spectrum at all scales, thus affecting the reconstruction of the cosmological parameters. Such interactions can lead to a slight increase of the value of $H_0$ and a slight decrease of $\sigma_8$, which can help reduce somewhat the tension between the CMB and lensing or Cepheids datasets. Here we show that it is impossible to solve both tensions simultaneously. While the 2015 Planck temperature and low multipole polarisation data combined with the Cepheids datasets prefer large values of the Hubble rate (up to $H_0 = 72.1^{+1.5}_{-1.7} \rm{km/s/Mpc}$, when $N_{\rm{eff}}$ is free to vary), the $\sigma_8$ parameter remains too large to reduce the $\sigma_8$ tension. Adding high multipole Planck polarization data does not help since this data shows a strong preference for low values of $H_0$, thus worsening current tensions, even though they also prefer smaller value of $\sigma_8$. 
\end{abstract}
\maketitle

\section{Introduction}
In the standard cosmological framework, dark matter is assumed to be collisionless. In practice this means that one arbitrarily sets the dark matter interactions  to zero when predicting the angular power spectrum of the Cosmic Microwave Background (CMB). However this treatment is at odds with the principle behind dark matter direct and indirect detection, where one explicitly assumes that dark matter (DM) interacts with ordinary matter. This is also in contradiction with the thermal hypothesis which relies on dark matter annihilations to explain the observed dark matter relic density. 

A more consistent approach consists in accounting for dark matter interactions and test whether they can be neglected by looking at their effects on cosmological observables. DM interactions in the early Universe damp the primordial dark matter fluctuations through the collisional damping mechanism \cite{deLaix:1995vi,Boehm:2000gq,Boehm:2004th}. They also affect the evolution of the other fluid(s) which the DM is interacting with. The two effects simultaneously impact the distribution of light and matter in the early Universe \cite{Boehm:2001hm} and eventually affect structure formation in the dark ages \cite{Boehm:2003xr}. They can also modify how our own cosmic neighborhood should look like  \cite{Schewtschenko:2015rno,Schewtschenko:2014fca,Boehm:2014vja,Vogelsberger:2015gpr,Cyr-Racine:2015ihg} and change the estimates of the cosmological parameters needed to account for the observed CMB anisotropies. 

The so-called "cut-off" scale at which one notices departures from the Lambda$+$Cold DM model (LCDM) predictions in the matter power spectrum is governed by the ratio of the elastic scattering cross section (corresponding to the dark matter scattering off the species $i$, normalised to the Thomson cross section $\sigma_T$) to the dark matter mass. We refer to this ratio as $$u_i = \frac{\sigma_{DM-i}}{\sigma_T} \ \left(\frac{m_{\rm{DM}}}{100 \rm{GeV}}\right)^{-1}.$$
The larger $u_i$, the higher the cut-off scale \cite{Boehm:2000gq,Boehm:2001hm,Boehm:2004th}. 

Dark matter-radiation interactions is the most interesting case among all interacting DM scenarios. Since radiation dominates the energy in the Universe for a very long time, such interactions erase the dark matter fluctuations on relatively large-scales for $u \ll 1 $ and also change the way the CMB looks like across the sky ~\cite{Boehm:2000gq,Boehm:2001hm,Chen:2002yh,Boehm:2004th,Sigurdson:2004zp,Mangano:2006mp,Serra:2009uu,CyrRacine:2012fz, Diamanti:2012tg, Diamanti:2012tg,Blennow:2012de,Dolgov:2013una,Wilkinson:2013kia,Dvorkin:2013cea}. 
Dark matter-baryon and dark matter self-interactions can also erase the DM fluctuations but the $u$ ratio needs to be of order 1 to produce the same effects as the one considered here~\cite{Boehm:2000gq,Boehm:2004th}, given that there  are many less baryons than radiation in the Universe and baryons are non-relativistic.

In what follows, we focus on Dark Matter-neutrino interactions and study their impact on the cosmological parameters (in particular the Hubble rate $H_0$, the effective number of relativistic degrees of freedom $\neff$ and the linear matter power spectrum value at 8 Mpc, $\sigma_8$). Previous analyses \cite{Wilkinson:2014ksa} indicated that Dark Matter-neutrino interactions prefer higher values of $H_0$ with respect to LCDM estimates. The higher $\neff$, the higher $H_0$. Therefore we investigate  whether DM-$\nu$ interactions could at least partially solve the current tensions arising between the CMB and late-time (i.e. strong lensing \cite{Bonvin:2016crt} and Cepheids \cite{R16}) measurements of the $H_0$ value. We also study whether DM-$\nu$ interactions could reduce the tension between the CMB-inferred value of $\sigma_8$ and large-scale-structure surveys,  owing to the damping they induce.   

In what follows, we consider the Planck 2015 data from the full mission duration, both the recommended Temperature plus low multipole polarisation information, as well as the complete spectral information, thereby including also the high multipole polarisation information which the Planck team considers as preliminary due to the presence of small but detectable low level residual systematics of ${\cal{O}}(1) \ \mu\rm{K}^2$  \cite{Aghanim:2015xee}. We briefly remind the reader of the expected impact of the DM interactions on the cosmological parameters in Section~\ref{sec:dmnurecap}. In Section~\ref{sec:method}, we present the method used to analyze the data and give the results in Sections~\ref{sec:temp}, \ref{sec:pol} and \ref{sec:r16}. We conclude in Section~\ref{sec:conclusion}.

\section{Impact of the DM-$\nu$ interactions on the cosmological parameters \label{sec:dmnurecap}}

\begin{figure}
\centering
\includegraphics[width=9cm]{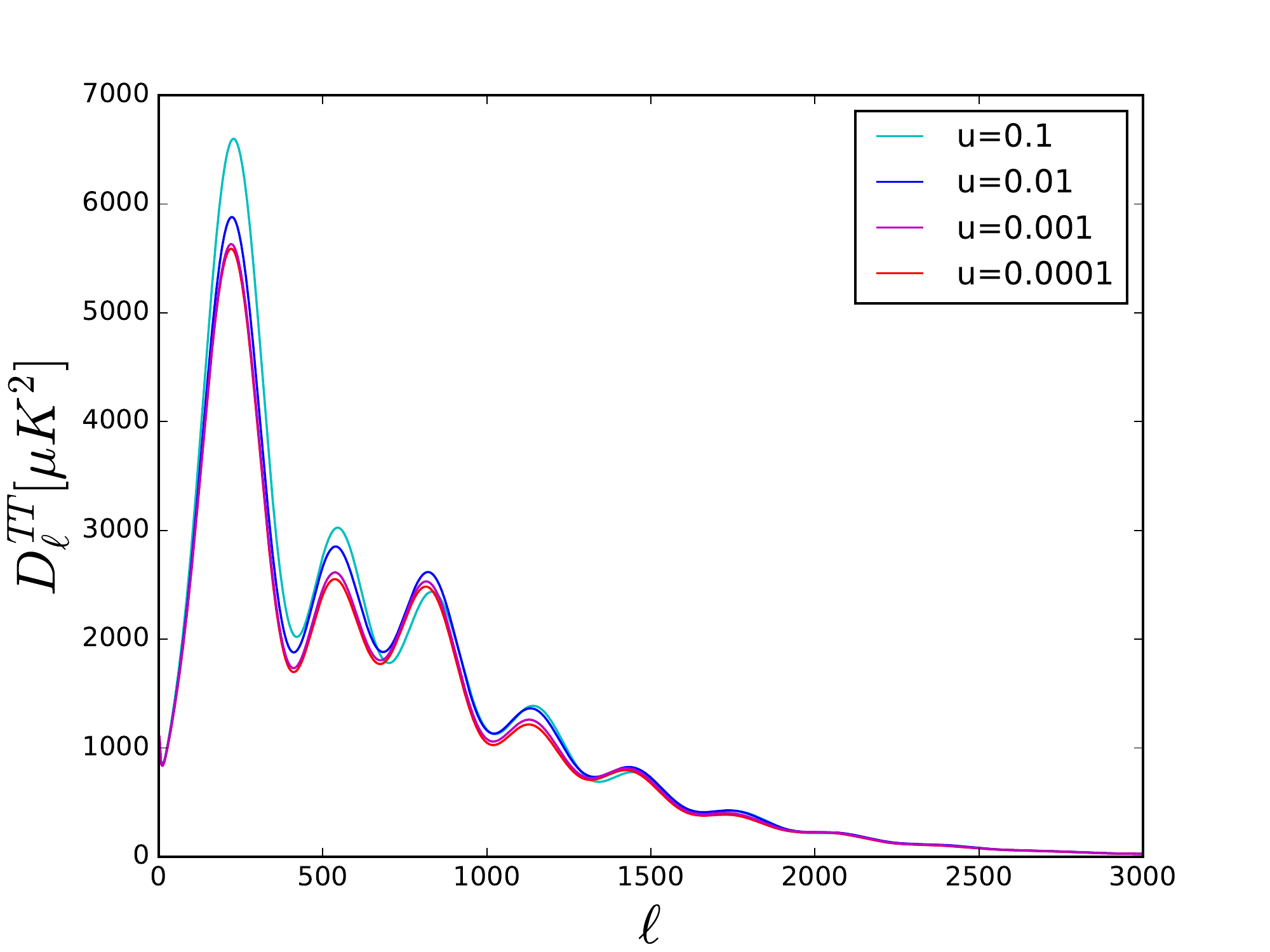}
\includegraphics[width=9cm]{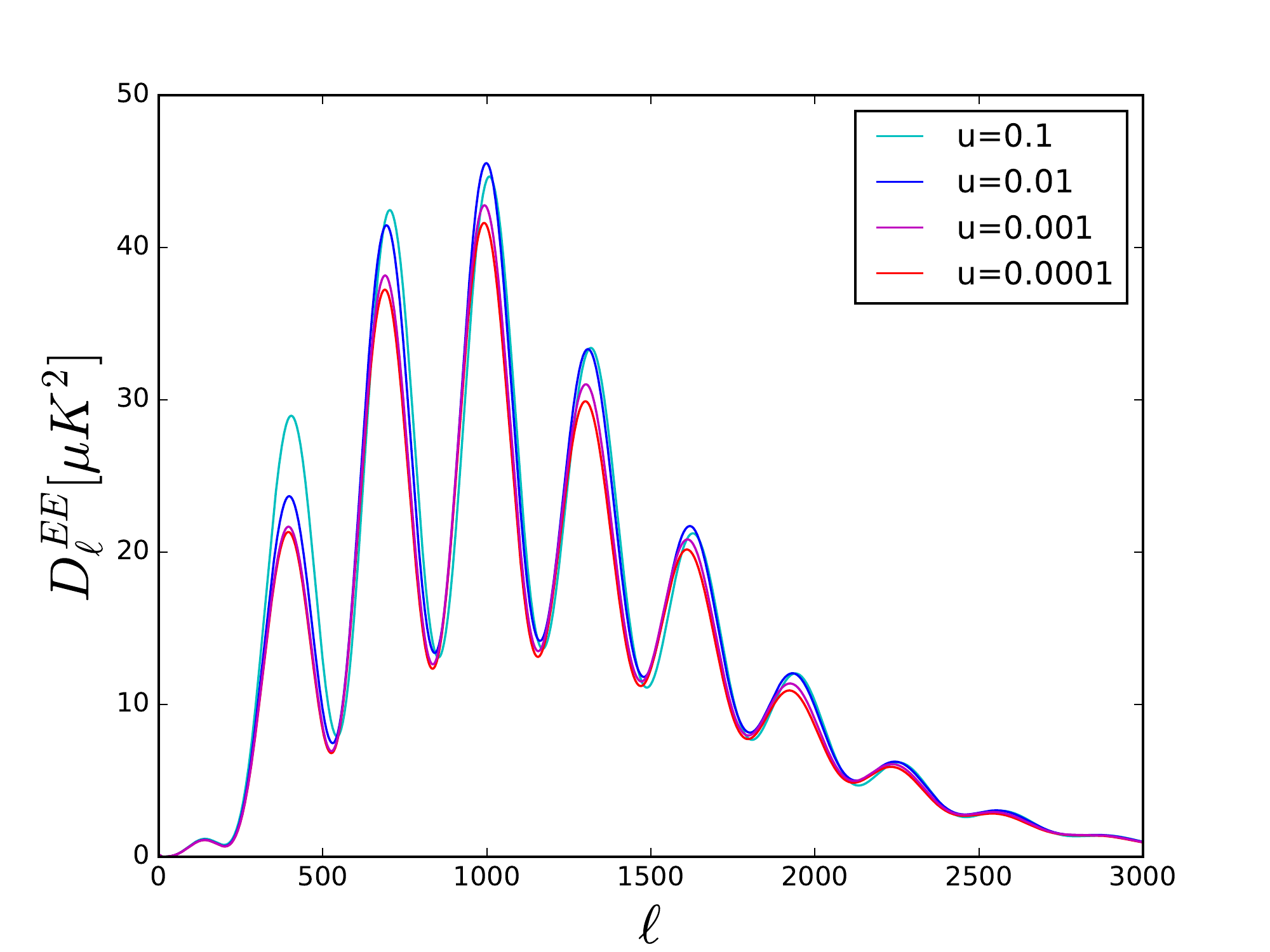}
\includegraphics[width=9cm]{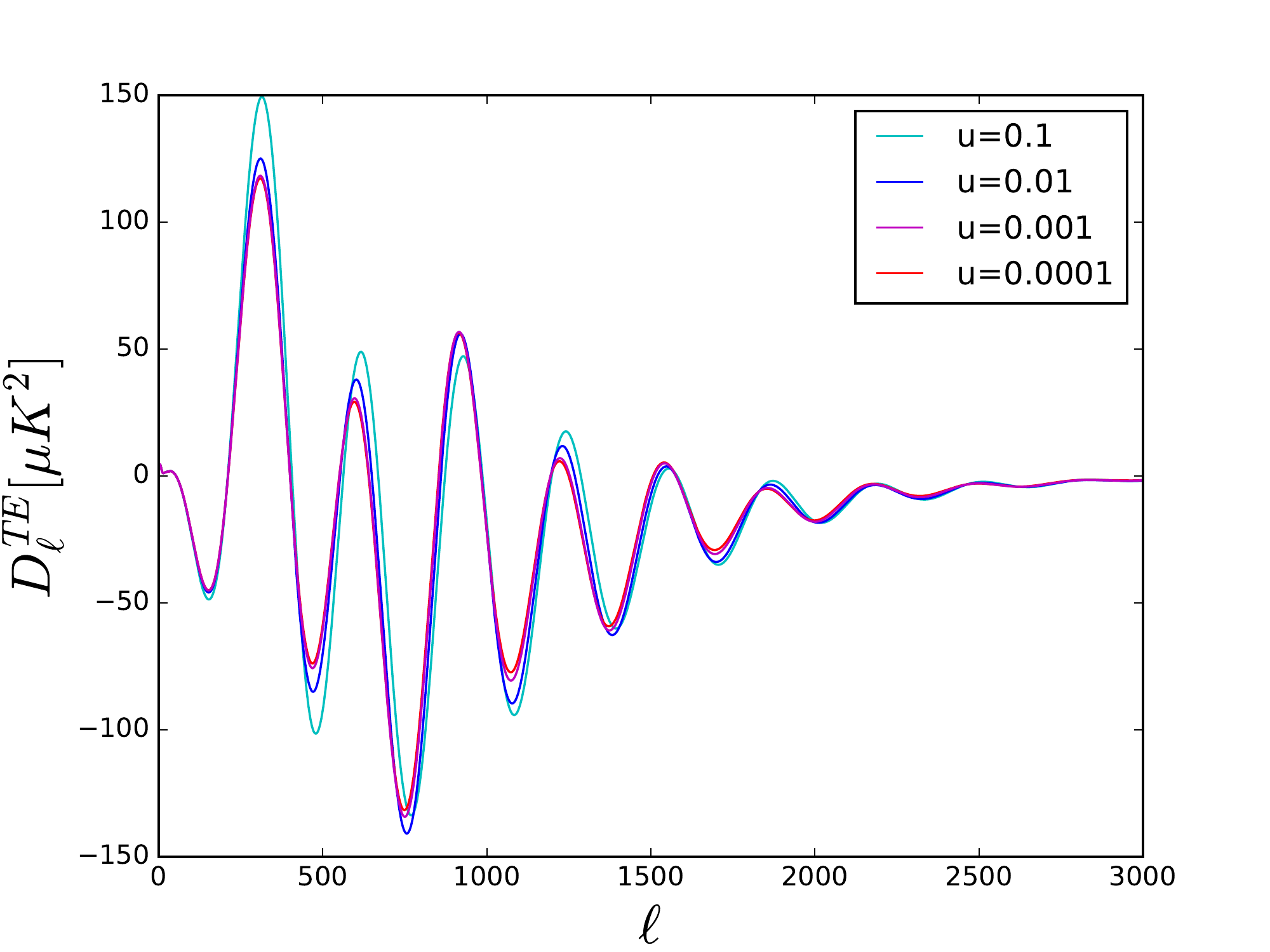}
\caption{The temperature and polarization CMB angular power spectra in the presence of Dark Matter-neutrino interactions.}
\label{fig1}
\end{figure}

The Dark Matter-neutrino interactions have five distinct effects on the temperature and polarisation angular power spectra. These were explained in Ref.~\cite{Wilkinson:2014ksa} and can be seen in Fig.~\ref{fig1}. 

Schematically, one can understand the impact of a DM-$\nu$ coupling on the cosmological parameters as follows. On one hand, the DM-$\nu$  interactions induce a damping of the DM fluctuations at small-scales (i.e. at high multipoles). On the other hand, they prevent the  neutrino free-streaming, till the neutrinos kinetically decouple from the DM. This last effect enhances the peaks at low multipoles, where the CMB temperature angular power spectrum is best measured. The greater the elastic scattering cross section (or the lighter the dark matter), the more pronounced are these two effects. Hence the fit to the data imposes an upper limit on the strength of these interactions.

The enhancement of the first few peaks is  less pronounced in a younger Universe. Hence scenarios with DM-$\nu$ interactions are  compatible with the data, when the value of $H_0$ is larger than the value estimated using the LCDM model.  One should  also observe a damping of the DM primordial fluctuations at small-scales because of the impact of neutrinos on the DM fluid. This effect translates into a damped oscillating matter power spectrum \cite{Boehm:2001hm} and thus leads to a smaller value of the $\sigma_8$ parameter than that in the LCDM scenario. 

Finally, we note that the difference in TE spectra between $u=10^{-3}$ and $u=10^{-4}$ is of the order of the same order of magnitude as Planck sensitivity (e.g. ${\cal{O}}(1) \  \mu\rm{K}^2$). Therefore Planck's angular power spectra alone are not sufficient to establish a preference for lower values of the $u$ ratio. However the suppression of power that such values ($u=10^{-3}$ and $u=10^{-4}$)  induce in the matter power spectrum are very different.  Using the $\sigma_8$ value together with the angular power spectra, we can rule out $u=10^{-3}$.

\section{Method \label{sec:method}}

The Boltzmann equations in presence of Dark Matter-neutrino interactions were given in e.g. Ref.~\cite{Mangano:2006mp, Serra:2009uu}.
To ensure the full treatment of the Boltzmann hierarchy,  we use a modified version of the Boltzmann code {\sc class}\footnote{\tt class-code.net}~\cite{Lesgourgues:2011re,Blas:2011rf}, that incorporates the Dark Matter-neutrino interactions \cite{Wilkinson:2014ksa}.
We perform our analysis in three main steps. 

In our first analysis, we use the six cosmological parameters of the Standard Model (namely the baryonic density $\Omega_bh^2$, the dark matter density $\Omega_ch^2$, the ratio between the sound horizon and the angular diameter distance at decoupling $\Theta_{s}$, the reionization optical depth $\tau$, the spectral index of the scalar perturbations $n_\mathrm{S}$, the amplitude of the primordial power spectrum $A_\mathrm{S}$) plus the ratio $u  \equiv u_{\nu}= \sigma_{DM-\nu}/m_{\rm{DM}}$.

In a second step, we consider eight free parameters, i.e the seven parameters mentioned above $+$ either the effective number of relativistic degrees of freedom $\neff$ or the total neutrino mass $\mnu$. Our rationale for doing this is that adding a dark radiation component ($\neff > 3.046$ \cite{std_neff}) as in Ref.~\cite{Wilkinson:2014ksa,darkradiation,DiValentino:2015sam} or allowing the sum of the neutrino masses $\mnu$ to depart from the benchmark value taken by the Planck collaboration ($\mnu = 0.06$ eV) could reduce current tensions on the age of the Universe.

Our last analysis uses nine free parameters, namely the seven mentioned above $+$ $\neff$  $+$ $\mnu$. Note that we use a logarithmic prior to constrain the $u$ parameter and  flat priors for the other parameters (i.e. $\Omega_bh^2$, $\Omega_ch^2$, $\Theta_{s}$, $\tau$, $n_\mathrm{S}$, $A_\mathrm{S}$, $\neff$  and $\mnu$).

To understand the impact of the polarisation data, we start by analysing
the full range of the 2015 temperature power spectrum ($2\leq\ell\leq2500$) plus the low multipoles polarization data ($2\leq\ell\leq29$) \cite{Aghanim:2015xee}. We will refer to this analysis as the ``Planck TT + lowTEB'' datasets.  We then perform a second analysis, which we will refer to as  ``Planck TTTEEE + lowTEB'',  where we include the Planck high multipole polarization data \cite{Aghanim:2015xee}. Finally, we perform a third analysis where we include the 2015 Planck measurements of the CMB lensing potential power spectrum $C^{\phi\phi}_\ell$ \cite{Ade:2015zua}. This last analysis will be referred to as the ``lensing'' dataset.

The scenarios for which the $H_0$ tension between the model-dependent Planck value and that inferred from the observations of Cepheids variables~\cite{R16} appears to be less than $2\sigma$ are analysed again. This time, we  assume a Gaussian prior on $H_0$ (i.e. $H_0=73.24\pm1.75 \ \rm{km \ s^{-1} \ Mpc^{-1}}$) and refer to this set of analysis as ``R16''.

\section{Results based on the ``Planck TT + lowTEB'' datasets only}\label{sec:temp}

\begin{table*}
\begin{center}\footnotesize
\scalebox{0.76}{\begin{tabular}{c|cccccccc}
  $\Lambda$CDM + $u$&  &  &  + $\neff$&  + $\neff$ &  + $\mnu$ &  + $\mnu$ &  + $\neff$ + $\mnu$&  + $\neff$ + $\mnu$  \\ 
\hline
Parameter         &  Planck TT & Planck TT & Planck TT & Planck TT & Planck TT  & Planck TT & Planck TT  & Planck TT  \\            
 &  + lowTEB &  + lowTEB + lensing &  + lowTEB& + lowTEB + lensing & + lowTEB & + lowTEB + lensing & + lowTEB &  + lowTEB + lensing  \\                 
\hline
\hspace{1mm}\\
$\Omega_bh^2$  &  $0.02224\,^{+0.00023}_{-0.00024}$& $0.02226\,^{+0.00027}_{-0.00026}$  & $0.02232\,^{+0.00037}_{-0.00041}$& $0.02234\,_{-0.00040}^{+0.00035}$ &  $0.02214\,^{+0.00027}_{-0.00026} $& $0.02217\,\pm 0.00029$  & $0.02219\,_{-0.00044}^{+0.00043}$& $0.02224\,_{-0.00042}^{+0.00037}$ \\
\hspace{1mm}\\
$\Omega_ch^2$  &  $0.1195\,_{-0.0023}^{+0.0022}$&  $0.1186\,_{-0.0022}^{+0.0021}$&  $0.1205\,^{+0.0039}_{-0.0045}$&  $0.1197\,_{-0.0041}^{+0.0039}$ &  $0.1200\,^{+0.0025}_{-0.0026}$&  $0.1195\,^{+0.0024}_{-0.0025}$&  $0.1206\,\pm 0.0046$&  $0.1206\,_{-0.0041}^{+0.0038}$\\
\hspace{1mm}\\
$\tau$ &   $0.079\,^{+0.018}_{-0.020}$& $0.070\,^{+0.015}_{-0.018}$&  $0.083\,_{-0.024}^{+0.018}$& $0.074\,_{-0.021}^{+0.016}$ &   $0.080\,_{-0.020}^{+0.018}$& $0.074\,^{+0.018}_{-0.019}$&  $0.083\,^{+0.021}_{-0.024}$& $0.077\,_{-0.022}^{+0.017}$ \\
\hspace{1mm}\\
$n_s$ &  $0.9652\,^{+0.0066}_{-0.0065}$&$0.9667\,_{-0.0065}^{+0.0071}$  & $0.969\,^{+0.015}_{-0.017}$&$0.971\,_{-0.017}^{+0.014}$ &  $0.9623\,^{+0.0083}_{-0.0082}$& $0.9640\,_{-0.0083}^{+0.0077}$  & $0.965\,\pm 0.018 $&$0.968\,_{-0.017}^{+0.015}$  \\
\hspace{1mm}\\
$ln(10^{10}A_s)$ & $3.091\,^{+0.034}_{-0.039}$& $3.071\,^{+0.027}_{-0.033}$& $3.100\,_{-0.053}^{+0.040}$& $3.080\,_{-0.044}^{+0.034}$ & $3.094\,^{+0.035}_{-0.039}$& $3.080\,^{+0.033}_{-0.034}$& $3.101\,^{+0.042}_{-0.054}$& $3.089\,_{-0.046}^{+0.036}$\\
\hspace{1mm}\\
$H_0 [\rm{Km \, s^{-1} \, Mpc^{-1}}]$ &$67.5\,\pm 1.0$ &$67.8\,\pm 1.0 $&$68.3\,^{+2.6}_{-3.2}$ &$68.7\,_{-3.0}^{+2.4}$ &$65.7\,^{+2.6}_{-1.9}$ &$66.2\,^{+2.2}_{-1.9}$&$66.2\,^{+4.0}_{-3.7}$ &$67.0\,_{-3.5}^{+3.3}$  \\
\hspace{1mm}\\
$\sigma_8$ &$0.825\,_{-0.016}^{+0.017}$& $0.814\,_{-0.012}^{+0.014} $  & $0.830\,_{-0.025}^{+0.021}$& $0.819\,_{-0.021}^{+0.019}$ &$0.788\,_{-0.033}^{+0.054}$& $0.787\,^{+0.036}_{-0.030}$  & $0.792\,_{-0.040}^{+0.060}$& $0.791\,_{-0.031}^{+0.041}$ \\
\hspace{1mm}\\
$u$  & $<-4.1$&$<-4.0$&$<-4.0$&$<-4.0$&$<-4.0 $&$<-4.1$& $<-4.0$&$<-4.0$\\
\hspace{1mm}\\
$\neff$ &$3.046$ &$3.046$& $3.14\,_{-0.35}^{+0.32}$ &$3.15\,_{-0.33}^{+0.28}$&$3.046$ &$3.046$ & $3.10\,\pm 0.35$ &$3.14\,_{-0.33}^{+0.30}$ \\
\hspace{1mm}\\
$\mnu [ \, eV]$ &$0.06$&$0.06$&$0.06$&$0.06$&$<2.0$&$<1.6$&$<2.2$& $<1.6$\\
\hline
\end{tabular}}
\end{center}
\caption{$68\%$~CL constraints on cosmological parameters with interactions, for the Planck TT + lowTEB and the Planck TT + lowTEB + lensing combination of datasets. When only upper limits are shown, they correspond to 95\% c.l. limits.}
\label{tab:temp}
\end{table*}

\begin{table*}
\begin{center}\footnotesize
\scalebox{0.74}{\begin{tabular}{c|cccccccc}
  $\Lambda$CDM&  &  &  + $\neff$&  + $\neff$ &  + $\mnu$ &  + $\mnu$ &  + $\neff$ + $\mnu$&  + $\neff$ + $\mnu$  \\ 
\hline
Parameter         & Planck TT & Planck TT & Planck TT& Planck TT & Planck TT  & Planck TT & Planck TT  & Planck TT  \\            
 & + lowTEB & + lowTEB + lensing&  + lowTEB& + lowTEB + lensing & + lowTEB& + lowTEB + lensing & + lowTEB & + lowTEB + lensing  \\                 
\hline
\hspace{1mm}\\
$\Omega_bh^2$  &  $0.02222\,\pm0.00023$& $0.02226\,\pm0.00023$  & $0.02230\pm0.00037$& $0.02232\,_{-0.00039}^{+0.00035}$ &  $0.02213\,\pm0.00027$& $0.02211\pm0.00026$  & $0.02215\,\pm0.00041$& $0.02212\pm0.00041$ \\
\hspace{1mm}\\
$\Omega_ch^2$  &  $0.1197\,\pm0.0022$&  $0.1186\,\,\pm0.0020$&  $0.1205\,\pm0.0041$&  $0.1195\,_{-0.0043}^{+0.0037}$ &  $0.1202\,\pm0.0024$&  $0.1199\,^{+0.0023}_{-0.0026}$&  $0.1205\,\pm0.0039$&  $0.1201\,\pm0.0039$\\
\hspace{1mm}\\
$\tau$ &   $0.078\,\pm0.019$& $0.066\,\pm0.016$&  $0.080\pm0.022$& $0.069\,\pm0.020$ &   $0.080\,\pm0.020$& $0.075\,\pm0.018$&  $0.081\,\pm0.021$& $0.076\,\pm0.20$ \\
\hspace{1mm}\\
$n_s$ &  $0.9655\,\pm0.0062$&$0.9677\,\pm0.0060$  & $0.969\,\pm0.016$&$0.971\,\pm0.015$ &  $0.9637\,\pm0.0071$& $0.9640\,\pm0.0068$  & $0.965\,\pm0.016$&$0.965\,\pm0.016$  \\
\hspace{1mm}\\
$ln(10^{10}A_s)$ & $3.089\,\pm0.036$& $3.062\,\pm0.029$& $3.096\,\pm0.047$& $3.070\,\pm0.042$ & $3.095\,\pm0.038$& $3.083\,\pm0.035$& $3.098\,\pm0.046$& $3.085\,\pm0.044$\\
\hspace{1mm}\\
$H_0 [\rm{Km \, s^{-1} \, Mpc^{-1}}]$ &$67.31\,\pm0.96$ &$67.81\,\pm0.92$&$68.0\,^{+2.6}_{-3.0}$ &$68.5\,_{-3.0}^{+2.5}$ &$65.6\,^{+3.1}_{-1.4}$ &$65.2\,^{+3.2}_{-2.0}$&$65.8\,^{+4.5}_{-3.3}$ &$65.3\,^{+4.2}_{-3.8}$  \\
\hspace{1mm}\\
$\sigma_8$ &$0.829\,\pm0.014$& $0.8149\,\pm0.0093$  & $0.834\,_{-0.025}^{+0.022}$& $0.820\,_{-0.021}^{+0.018}$ &$0.796\,_{-0.023}^{+0.057}$& $0.776\,^{+0.047}_{-0.025}$  & $0.796\,_{-0.030}^{+0.065}$& $0.777\,_{-0.035}^{+0.052}$ \\
\hspace{1mm}\\
$\neff$ &         $3.046$ &$3.046$& $3.13\,_{-0.34}^{+0.30}$ &$3.13\,_{-0.34}^{+0.29}$&         $3.046$ &$3.046$ & $3.08\,\pm 0.31$ &$3.07\,\pm0.31$ \\
\hspace{1mm}\\
$\mnu [ \, eV]$ &              $0.06$&$0.06$&$0.06$&$0.06$&$<0.715$&$<0.675$&$<0.725$& $<0.677$\\
\hline
\end{tabular}}
\end{center}
\caption{$68\%$~CL constraints on cosmological parameters without interactions, for the Planck TT + lowTEB and the Planck TT + lowTEB + lensing combination of datasets. When only upper limits are shown, they correspond to 95\% c.l. limits.}
\label{tab:temp_nou}
\end{table*}

We now present the results of our analyses using the Planck low multipole polarisation data.  The $68 \% $ confidence level (c.l.)  limits on the cosmological parameters for the DM-$\nu$ scenario are shown in Table~\ref{tab:temp}. For comparison, we also display the $68 \% $ c.l.  constraints obtained by the Planck collaboration \cite{planck2015} for collisionless LCDM\footnote{\url{https://wiki.cosmos.esa.int/planckpla2015/images/f/f7/Baseline_params_table_2015_limit68.pdf}} in Table~\ref{tab:temp_nou}.

 ''Weak'' interactions are expected to erase primordial scales which have not been observed yet. Hence our analysis is bound to exclude  only the strongest DM-$\nu$ interactions. This translates into an upper bound on the $u$ parameter of $u < 10^{-4.1}$ (or $u<10^{-4.0}$, using the lensing datasets),  corresponding to a DM-$\nu$ elastic cross section of $\sigma \simeq 3-6 \ 10^{-31} \ \left(m_{\rm{dm}}/\rm{GeV}\right) \rm{cm^2}$. This result is similar to the limit derived in Ref.~\cite{Escudero:2015yka}, using the 2013 Planck temperature data. Furthermore, we find that the Planck data prefer  low values of the Hubble constant $H_0$, even in the presence of DM interactions (see Table.~\ref{tab:temp}). This is at odds with the conclusions from Ref.~\cite{Wilkinson:2013kia} but a possible explanation is that the 2013 Planck data relied on the (low $l$) WMAP polarisation data, while the 2015 Planck data rely on Planck's  polarisation data.

We observe in addition that the introduction of the DM-$\nu$ interactions breaks the well-known degeneracy between $H_0$ and the clustering parameter $\sigma_8$. The Hubble constant slightly increases while the  clustering parameter $\sigma_8$ slightly decreases in presence of such interactions. 
For example, we find $\sigma_8=0.825\, _{-0.015}^{+0.014}$ (see the first column of the Table~\ref{tab:temp}) while the Planck collaboration found $\sigma_8=0.829\,\pm 0.014$  using 
the same dataset combination (see the first column of the Table~\ref{tab:temp_nou}) for collisionless LCDM.

When we allow $\neff$ to vary, we obtain $\neff=3.14\,_{-0.35}^{+0.32}$ for $\mnu =0.06$ eV (see the third column of the Table~\ref{tab:temp} and Fig.~\ref{fig2}). This result is a bit higher than the Standard Model value ($\neff = 3.046$) but it does remain compatible with it nonetheless. The Hubble rate then shifts from $H_0 = 68.0\,^{+2.6}_{-3.0} \ \rm{km \ s^{-1} \ Mpc^{-1} }$ to $H_0 = 68.3^{+2.6}_{-3.2} \ \rm{km \ s^{-1} \ Mpc^{-1}}$ (see the third columns of  Table~\ref{tab:temp_nou} and Table~\ref{tab:temp} respectively). In this case, the tension between the local measurements of $H_0$ ($H_0=73.24 \pm 1.75 \ \rm{km \ s^{-1} \ Mpc^{-1}}$) \cite{R16}, and the Planck $\Lambda$CDM value \cite{planck2015} is somewhat reduced. Therefore we can reasonably combine the Planck datasets with the R16 datasets and perform a new analysis. The results are given in Table \ref{tab:R16}. 

The introduction of DM-$\nu$ interactions is also compatible with  heavier neutrinos. This is an important point since it was noted in Ref.~\cite{Giusarma:2014zza,sigma8,DiValentino:2015sam} that massive neutrinos could alleviate the tension between Planck and the weak lensing measurements from the CFHTLenS survey \cite{Heymans:2012gg, Erben:2012zw} and KiDS-450 \cite{Hildebrandt:2016iqg}.  Assuming DM-$\nu$ interactions and the Planck TT + lowTEB + lensing dataset, we obtain $\mnu < 1.6 \ \rm{eV}$ at $95\%$~ c.l.  (see the sixth column of  Table~\ref{tab:temp}) instead of $\mnu<0.675 \ \rm{eV}$ for LCDM (see the sixth column of Table~\ref{tab:temp_nou}). 

For that same combination of datasets (Planck TT + lowTEB + lensing), both the Hubble constant $H_0$ and the clustering parameter $\sigma_8$ increase with respect to the Standard Model ($\Lambda$CDM + $\mnu$) value. However, the value of $\sigma_8$ thus obtained remains small enough to partially reduce the tension with the weak lensing measurements. We obtain $\sigma_8 = 0.787\,^{+0.036}_{-0.030}$ which is much lower than the Planck value  $\sigma_8= 0.8149\,\pm0.0093$, which was reported by the collaboration for LCDM  only (i.e LCDM $+$ fixed values of $\neff$ and $\mnu$) using the Planck TT + lowTEB + lensing dataset. 
Using the Planck TT+lowTEB dataset only and the definition $S_8 \equiv\sigma_8 \sqrt{\Omega_m/0.3}$, we find  $S_8 =0.826\, _{-0.028}^{+0.033}$. Hence adding the DM-$\nu$ interactions does reduce the tension with the KiDS-450 measurements ($S_8=0.745\pm0.039$ \cite{Hildebrandt:2016iqg}), to about $1.7\sigma$.

Finally, varying $\neff$ and $\mnu$ simultaneously allows to reduce the $H_0$ and $\sigma_8$ tensions. Using the Planck TT + lowTEB datasets, we find that $H_0=66.2\, _{-3.7}^{+4.0}$ and $\sigma_8= 0.792\,_{-0.040}^{+0.060}$, as shown in the seventh column of the Table~\ref{tab:temp}.
The tension with other $H_0$ measurements is then about $1.6\sigma$. The new value for $S_8$ (namely $S_8=0.826\, _{-0.027}^{+0.033}$) also reduces the tension with KiDS-450 \cite{Hildebrandt:2016iqg} to  about $1.7\sigma$.

\section{Results with the polarization data}\label{sec:pol}

\begin{table*}
\begin{center}\footnotesize
\scalebox{0.74}{\begin{tabular}{c|cccccccc}
  $\Lambda$CDM + $u$&  &  &  + $\neff$&  + $\neff$ &  + $\mnu$ &  + $\mnu$ &  + $\neff$ + $\mnu$&  + $\neff$ + $\mnu$  \\ 
\hline
Parameter         & Planck TTTEEE & Planck TTTEEE & Planck TTTEEE & Planck TTTEEE & Planck TTTEEE & Planck TTTEEE & Planck TTTEEE & Planck TTTEEE   \\            
 & + lowTEB & + lowTEB + lensing & + lowTEB& + lowTEB + lensing & + lowTEB & + lowTEB + lensing  & + lowTEB& + lowTEB + lensing  \\                    
\hline
\hspace{1mm}\\
$\Omega_bh^2$  &  $0.02225\,\pm 0.00017$& $0.02225,^{+0.00017}_{-0.00018}$  & $0.02218\,\pm 0.00028$& $0.02216\, ^{+0.00023}_{-0.00025}$ &  $0.02219\,^{+0.00018}_{-0.00017}$& $0.02219\,\pm 0.00018$  & $0.02212\, ^{+0.00029}_{-0.00031}$& $0.02210\,^{+0.00025}_{-0.00026}$ \\
\hspace{1mm}\\
$\Omega_ch^2$  &  $0.1198\,^{+0.0016}_{-0.0015}$&  $0.1194\,\pm 0.0015$&  $0.1190\,_{-0.0036}^{+0.0035}$&  $0.1179\,\pm 0.0030$ &  $0.1200\,\pm 0.0016 $&  $0.1197\,^{+0.0015}_{-0.0016}$&  $0.1188\,_{-0.0037}^{+0.0038}$&  $0.1185\,\pm 0.0032$\\
\hspace{1mm}\\
$\tau$ &   $0.080\,^{+0.016}_{-0.018}$& $0.066\,^{+0.013}_{-0.015}$&  $0.078\,^{+0.018}_{-0.019}$& $0.065\, ^{+0.011}_{-0.015}$ &   $0.082\, _{-0.017}^{+0.018}$& $0.073\,^{+0.015}_{-0.016}$&  $0.080\, ^{+0.019}_{-0.021}$& $0.071\, _{-0.016}^{+0.014}$ \\
\hspace{1mm}\\
$n_s$ &  $0.9639\,^{+0.0053}_{-0.0052}$&$0.9644\,^{+0.0056}_{-0.0054}$  & $0.961\,\pm 0.011$&$0.9603\,^{+0.0093}_{-0.0095}$ &  $0.9620\,^{+0.0060}_{-0.0056}$&$0.9628\,_{-0.0055}^{+0.0057}$  & $0.959\,_{-0.013}^{+0.012} $&$0.959\,\pm 0.010$  \\
\hspace{1mm}\\
$ln(10^{10}A_s)$ & $3.093\,^{+0.032}_{-0.035}$& $3.065\,^{+0.024}_{-0.027}$& $3.087\,_{-0.041}^{+0.040}$& $3.059\,^{+0.024}_{-0.030}$ & $3.099\,^{+0.035}_{-0.033}$& $3.079\,^{+0.028}_{-0.031}$& $3.092\, _{-0.043}^{+0.041}$& $3.072\,^{+0.030}_{-0.034}$\\
\hspace{1mm}\\
$H_0 [\rm{Km \, s^{-1} \, Mpc^{-1}}]$ &$67.32\,_{-0.71}^{+0.70}$ &$67.50\,^{+0.70}_{-0.71}$&$66.8\,^{+1.8}_{-1.9}$ &$66.8\,\pm 1.6$ &$66.0\,^{+2.3}_{-1.2}$ &$66.1\,^{+1.9}_{-1.3}$&$65.4\, ^{+2.8}_{-2.5}$ &$65.4\,^{+2.2}_{-2.0}$  \\
\hspace{1mm}\\
$\sigma_8$ &$0.827\,^{+0.016}_{-0.015}$& $0.814\,^{+0.013}_{-0.012}$  & $0.822\,\pm 0.023$& $0.809\,^{+0.013}_{-0.014}$ &$0.797\,_{-0.023}^{+0.049}$& $0.789\,^{+0.036}_{-0.020}$  & $0.791\,_{-0.050}^{+0.052}$& $0.784\,^{+0.035}_{-0.024}$ \\
\hspace{1mm}\\
$u$  & $<-4.1$&$<-4.1$&$<-4.0$&$<-4.0$  & $<-4.1 $&$<-4.2$&            $<-3.9$&$<-4.3$\\
\hspace{1mm}\\
$\neff$ &         $3.046$ &$3.046$& $2.98\,^{+0.23}_{-0.24}$ &$2.94\,\pm 0.20$&         $3.046$ &$3.046$ & $2.96\,^{+0.23}_{-0.28}$ &$2.95\,^{+0.20}_{-0.21}$ \\
\hspace{1mm}\\
$\mnu [\, eV]$ &              $0.06$&$0.06$&$0.06$&$0.06$&$<1.9$&$<1.5$&$<2.0$&$<1.6$ \\
\hline
\end{tabular}}
\end{center}
\caption{$68\%$~CL constraints on cosmological parameters with interactions, for the Planck TTTEEE + lowTEB and the Planck TTTEEE + lowTEB + lensing combination of datasets. When only upper limits are shown, they correspond to 95\% c.l. limits.}
\label{tab:pol}
\end{table*}

\begin{table*}
\begin{center}\footnotesize
\scalebox{0.74}{\begin{tabular}{c|cccccccc}
 $\Lambda$CDM &   &  & + $\neff$&  + $\neff$ & + $\mnu$ &  + $\mnu$ &  + $\neff$ + $\mnu$&  + $\neff$ + $\mnu$  \\ 
\hline
Parameter         & Planck TTTEEE & Planck TTTEEE & Planck TTTEEE & Planck TTTEEE & Planck TTTEEE & Planck TTTEEE& Planck TTTEEE & Planck TTTEEE   \\            
 & + lowTEB & + lowTEB + lensing & + lowTEB& + lowTEB + lensing & + lowTEB & + lowTEB + lensing & + lowTEB& + lowTEB + lensing  \\  
 \hline
\hspace{1mm}\\
$\Omega_bh^2$  &  $0.02225\,\pm0.00016$& $0.02226\pm0.00016$  & $0.02220\, \pm0.00024$& $0.02216\pm0.00023$ &  $0.02222\,\pm0.00017$& $0.02219\pm0.00017$  & $0.02215\, \pm0.00025$& $0.02208\pm0.000025$ \\
\hspace{1mm}\\
$\Omega_ch^2$  &  $0.1198\,\pm0.0015$&  $0.1193\,\pm0.0014$&  $0.1191\,\pm0.0031$&  $0.1178\,\pm0.0030$ &  $0.1200\,\pm0.0015$&  $0.1198\,\pm0.0015$&  $0.1191\,\pm0.0031$&  $0.1184\,\pm0.0030$\\
\hspace{1mm}\\
$\tau$ &   $0.079\,\pm0.017$& $0.063\,\pm0.014$&  $0.077\pm0.0018$& $0.060\, \pm0.014$ &   $0.083\, \pm 0.018$& $0.074\,\pm0.017$&  $0.081\, \pm0.018$& $0.071\, \pm0.018$ \\
\hspace{1mm}\\
$n_s$ &  $0.9645\,\pm0.0049$&$0.9653\,\pm0.0048$  & $0.9620\,\pm0.0097$&$0.9606\,\pm0.0092$ &  $0.9639\,\pm0.0050$&$0.9637\,\pm0.0051$  & $0.9610\,\pm0.0099$&$0.9589\,\pm0.0095$  \\
\hspace{1mm}\\
$ln(10^{10}A_s)$ & $3.094\,\pm0.034$& $3.059\,\pm0.025$& $3.088\,\pm0.038$& $3.049\,\pm0.029$ & $3.100\,\pm0.034$& $3.081\,\pm0.033$& $3.095\,\pm0.039$& $3.071\,\pm0.037$\\
\hspace{1mm}\\
$H_0 [\rm{Km \, s^{-1} \, Mpc^{-1}}]$ &$67.27\,\pm 0.66$ &$67.51\,\pm0.64$&$66.8\,\pm1.6$ &$66.7\,\pm1.5$ &$66.3\,^{+2.0}_{-0.9}$ &$65.6\,^{+2.5}_{-1.4}$&$65.8\, ^{+2.6}_{-1.8}$ &$64.8\,^{+2.5}_{-2.1}$  \\
\hspace{1mm}\\
$\sigma_8$ &$0.831\,\pm0.013$& $0.8150\,\pm0.0087$  & $0.828\,\pm0.018$& $0.809\,\pm0.013$ &$0.812\,_{-0.017}^{+0.039}$& $0.783\,^{+0.040}_{-0.020}$  & $0.807\,_{-0.044}^{+0.022}$& $0.778\,^{+0.038}_{-0.024}$ \\
\hspace{1mm}\\
$\neff$ &         $3.046$ &$3.046$& $2.99\,\pm0.20$ &$2.94\,\pm 0.20$&         $3.046$ &$3.046$ & $2.98\,\pm0.20$ &$2.93\,\pm 0.19$ \\
\hspace{1mm}\\
$\mnu [\, eV]$ &              $0.06$&$0.06$&$0.06$&$0.06$&$<0.492$&$<0.589$&$<0.494$&$<0.577$ \\
\hline
\end{tabular}}
\end{center}
\caption{$68\%$~CL constraints on cosmological parameters without interactions, for the Planck TTTEEE + lowTEB and the Planck TTTEEE + lowTEB + lensing combination of datasets. When only upper limits are shown, they correspond to 95\% c.l. limits.}
\label{tab:pol_nou}
\end{table*}

In Table~\ref{tab:pol}, we  report the $68 \% $ c.l  limits  on the DM-$\nu$ scenario obtained using the polarisation data. 
For comparison, we also give the $68 \% $ c.l.  limits\footnote{\url{https://wiki.cosmos.esa.int/planckpla2015/images/f/f7/Baseline_params_table_2015_limit68.pdf}}  
obtained by the Planck collaboration \cite{planck2015} for the LCDM scenario in Table~\ref{tab:pol_nou}.  

Assuming fixed values of $\neff$ and $\mnu$, we find that the use of the Planck polarization data generally slightly improves the constraints of the strength of the Dark Matter-neutrino interactions (see Fig.~\ref{fig3}). For example, instead of $u<10^{-4.0}$, we now find  $u< 10^{-4.3}$ using the Planck TTTEEE + lowTEB + lensing datasets and the scenario with nine parameters (see last column of Table~\ref{tab:pol}). The  rest of the parameters remain compatible with $\Lambda$CDM values.

Similarly to the analysis performed in Section~\ref{sec:temp}, we also vary the effective number of relativistic degrees of freedom $\neff$. However it remains consistent with the standard model value, and so does the Hubble constant $H_0$ in this case (see, for example, the third column of the Table~\ref{tab:pol} and~\ref{tab:pol_nou}). 

Adding the polarization data however helps to relax the bounds on massive neutrinos. The latter shifts from $\mnu<0.492 \ \rm{eV}$ for the Planck TTTEEE + lowTEB datasets without interactions  to $\mnu<1.9 \ \rm{eV}$ for the same combination of datasets in presence of interactions (see the fifth column of  Table~\ref{tab:pol_nou} and Table~\ref{tab:pol} respectively). Furthermore, $\sigma_8$ decreases a bit in presence of interactions. We find $\sigma_8= 0.797\,_{-0.023}^{+0.049}$ in presence of interactions ($\sigma_8= 0.789\,_{-0.020}^{+0.036}$ if we add the lensing dataset) versus $\sigma_8= 0.812\,^{+0.039}_{-0.017}$ (or $\sigma_8= 0.783\,_{-0.020}^{+0.040}$ if we add the lensing dataset) in LCDM, as shown in the fifth and six columns of Tables~\ref{tab:pol} and ~\ref{tab:pol_nou}.
Here again, we find that the tension with the weak lensing measurements is reduced. We obtain $S_8=0.832\, _{-0.022}^{+0.029}$ in presence of interactions (and letting $\mnu$ free to vary) for the Planck TTTEEE+lowTEB datasets, which decreases the tension with KiDS-450 to $1.8\sigma$.

Finally, we observe a small shift in both values of $H_0$ and $\sigma_8$ with respect to LCDM when we either vary $\neff$, $\mnu$ or both simultaneously. The upper bound on $\mnu$ is also relaxed with respect to the collisionless $\Lambda$CDM. Adding the lensing dataset, we obtain  $H_0=65.4\, ^{+2.2}_{-2.0}$ and $\sigma_8= 0.784\,_{-0.024}^{+0.035}$ ($S_8=0.820_{-0.015}^{+0.019}$), as shown in the eight column of the Table~\ref{tab:pol}. Whilst the new value of $H_0$ does not remove completely the tensions between the different  observation datasets, it does reduce the $S_8$ tension to $1.7\sigma$. 

We note that when the Dark Matter-neutrino interactions are introduced, the scalar spectral index ($n_S$) gets very slightly shifted towards smaller values, for all the dataset combinations and parameters considered in this paper. This shift is due to the fact that the interactions change all the acoustic peaks (see Fig.~\ref{fig1} and discussion in Section II). In fact, they increase the low multipoles due to the suppression of neutrino free-streaming and decrease the high multipoles due to the collisional damping. Therefore, in order to reconcile the prediction of this model with the observed angular power spectra, the increase in the Hubble constant needs to be compensated by a change in the spectrum tilt.

\section{Results with R16}\label{sec:r16}

\begin{table*}
\begin{center}\footnotesize
\scalebox{0.76}{\begin{tabular}{c|cc}
  $\Lambda$CDM + $u$  &   + $\neff$ &    + $\neff$ + $\mnu$  \\ 
\hline
Parameter         &  Planck TT &  Planck TT  \\            
 &  + lowTEB + R16&   + lowTEB + R16  \\             
\hline
\hspace{1mm}\\
$\Omega_bh^2$  & $0.02278\,^{+0.00026}_{-0.00025}$& $0.02278\,\pm 0.00027$  \\
\hspace{1mm}\\
$\Omega_ch^2$  &  $0.1238\,^{+0.0037}_{-0.0038}$&  $0.1240\,_{-0.0045}^{+0.0035}$\\
\hspace{1mm}\\
$\tau$  &     $0.099\,^{+0.019}_{-0.021}$& $0.100\,_{-0.021}^{+0.023}$ \\
\hspace{1mm}\\
$n_s$  & $0.9898\,_{-0.0094}^{+0.0088} $&$0.990\,_{-0.010}^{+0.009}$  \\
\hspace{1mm}\\
$ln(10^{10}A_s)$ & $3.143\,^{+0.041}_{-0.039}$& $3.145\,_{-0.037}^{+0.054}$\\
\hspace{1mm}\\
$H_0 [\rm{Km \, s^{-1} \, Mpc^{-1}}]$ &$72.1\,^{+1.5}_{-1.7}$ &$71.9\,_{-1.8}^{+1.6}$  \\
\hspace{1mm}\\
$\sigma_8$   & $0.850\,_{-0.018}^{+0.024}$& $0.846\,_{-0.025}^{+0.030}$ \\
\hspace{1mm}\\
$u$   & $<-4.0$&$<-4.0$\\
\hspace{1mm}\\
$\neff$ &         $3.54\,\pm 0.20$ &$3.56\,_{-0.26}^{+0.19}$ \\
\hspace{1mm}\\
$\mnu [ \, eV]$ &$0.06$& $<0.87$\\
\hline
\end{tabular}}
\end{center}
\caption{$68\%$~CL constraints on cosmological parameters with interactions, for the Planck TT + lowTEB + R16 combination of datasets. If only upper limits are shown, they are at 95\% c.l.}
\label{tab:R16}
\end{table*}

In this Section, we analyse again the models for which the tension between the 2015 Planck and  Riess et al. 2016  \cite{R16} value of $H_0$ 
is less than $2\sigma$. These correspond to the scenarios where $\neff$ is free to vary, when we ignored the high multipole polarisation data. Applying a Gaussian prior on the value of $H_0$, we obtain new constraints on the cosmological parameters ($68 \% $ C.L.) for the interacting DM scenario, as shown in Table~\ref{tab:R16}.

\begin{figure*}
\centering
\includegraphics[width=1.5\columnwidth]{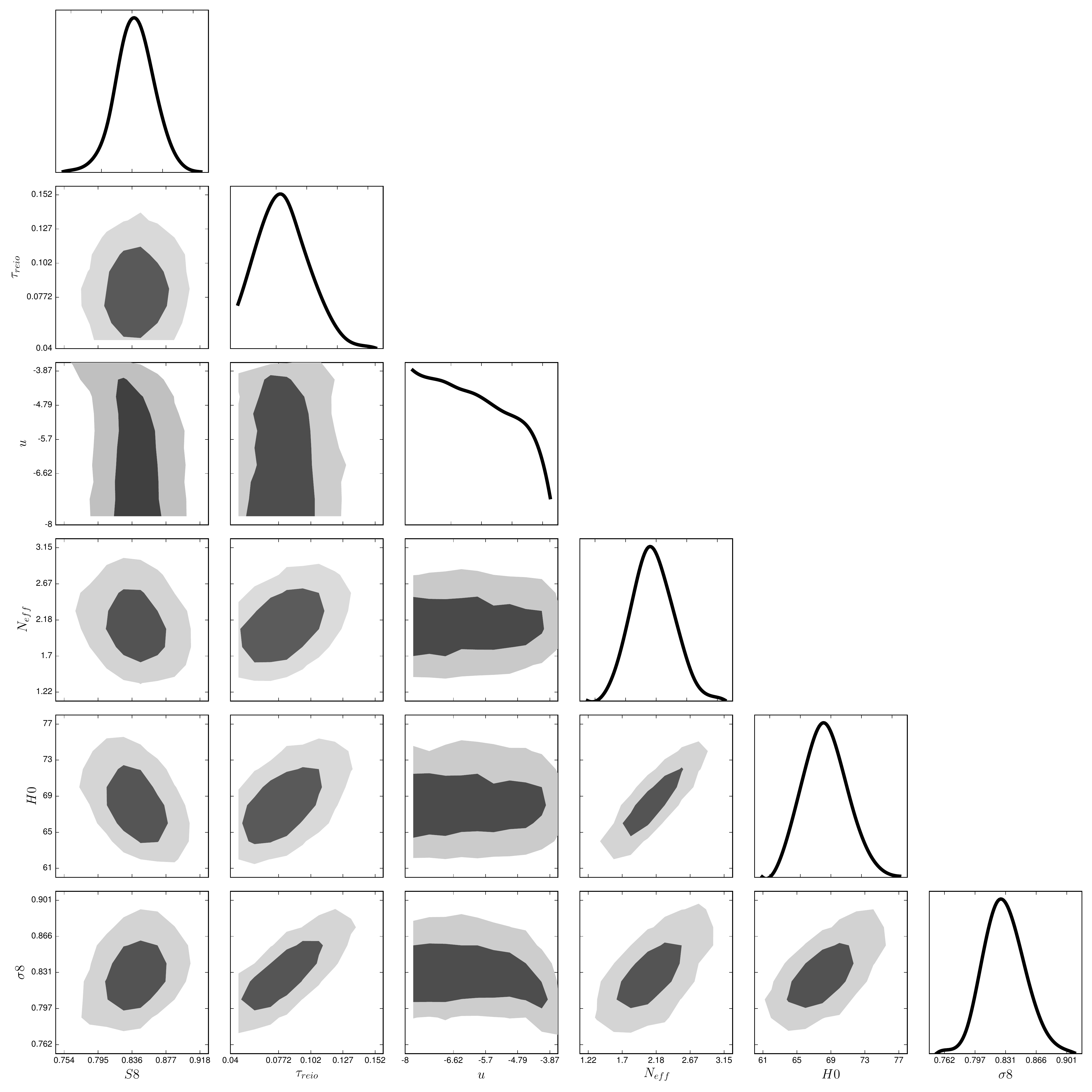}
\caption{Triangle plot showing the 1D and 2D posterior distributions of the cosmological parameters for Planck TT + lowTEB in the $\Lambda$CDM + $u$ + $\neff$ scenario.}
\label{fig2}
\end{figure*}

\begin{figure*}
\centering
\includegraphics[width=1.5\columnwidth]{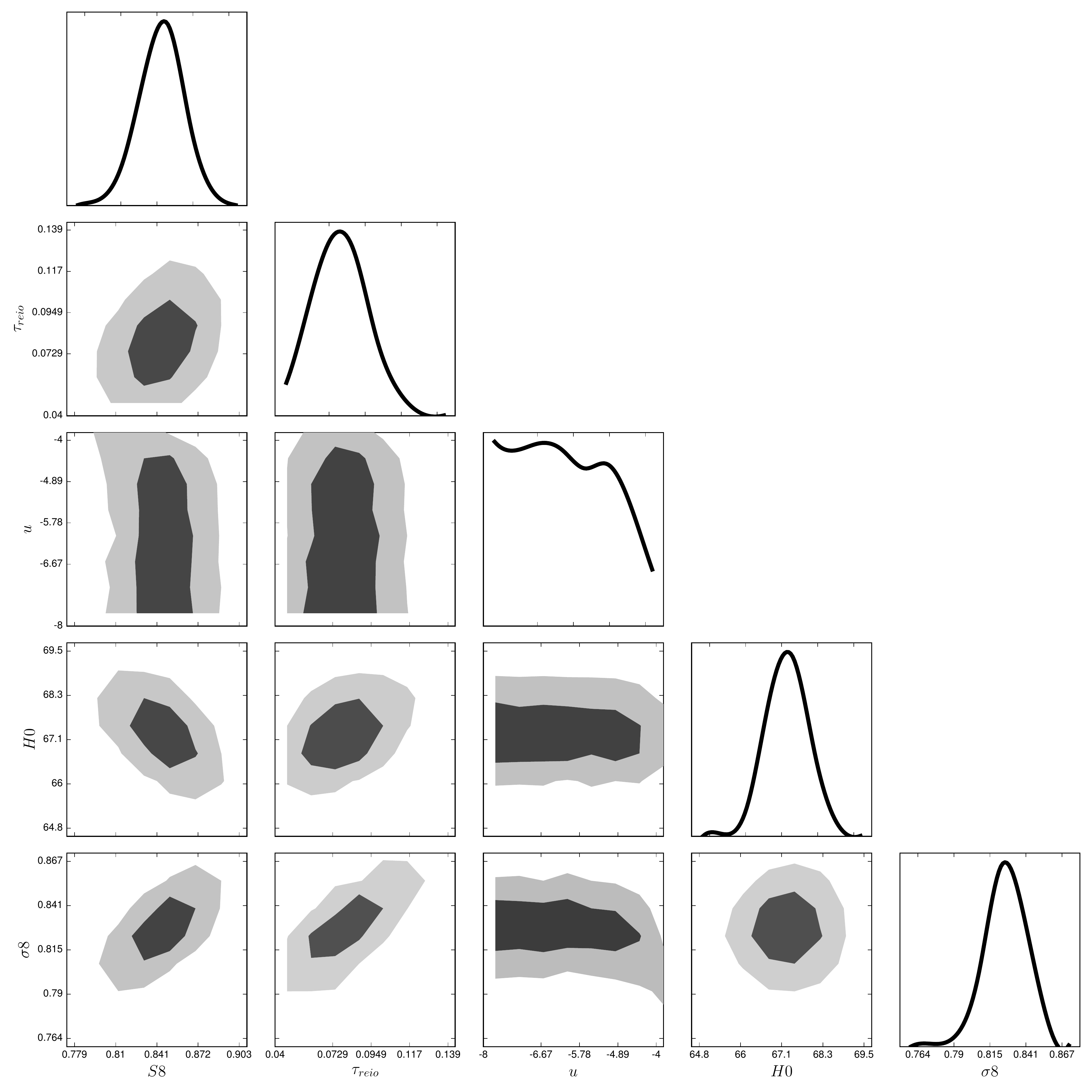}
\caption{Triangle plot showing the 1D and 2D posterior distributions of the cosmological parameters for Planck TTTEEE + lowTEB in the $\Lambda$CDM + $u$ scenario.}
\label{fig3}
\end{figure*}

We find that all the cosmological parameters are shifted towards higher values, as can be seen by comparing the results from  Table~\ref{tab:R16} with Table~\ref{tab:temp}. Moreover, owing to the very well-known degeneracy between $H_0$ and $\neff$ (see Fig.~\ref{fig2}), we find an indication for a dark radiation at about $2\sigma$ by imposing the R16 prior. In particular, we find $\neff=3.54\,\pm 0.20$ for the $\Lambda$CDM + $u$ + $\neff$ scenario and $\neff=3.56\,_{-0.26}^{+0.19}$ for the $\Lambda$CDM + $u$ + $\neff$ + $\mnu$ model. A dark radiation component can be explained by the existence of some extra relic component, such as a sterile neutrino or a thermal axion \cite{darkradiation, DiValentino:2015wba,DiValentino:2016ikp, Giusarma:2014zza, Archidiacono:2011gq}. However, in these models, an increase in the value  of $\neff$ may not be related to the presence of an additional species. It could be related to dark matter annihilations into neutrinos as they would reheat the neutrino fluid and mimic an increase in the value of $\neff$ \cite{Boehm:2012gr,Boehm:2013jpa}.

\section{Conclusion \label{sec:conclusion}}

In the $\Lambda$CDM model, dark matter is assumed to be collisionless. This means that one arbitrarily sets the dark matter interactions to zero to interpret the CMB temperature and polarisation angular power spectra and determine the cosmological parameters. Here we relaxed the collisionless  assumption and studied  the impact of DM-$\nu$ interactions on the cosmological parameters. 

We performed a similar analysis to \cite{Wilkinson:2014ksa}. However this time, we used the full 2015 Planck  data \cite{planck2015} as they include both the high and low multipoles polarization spectra and are more precise than the 2013 data. In general, we observe that the introduction of dark matter-neutrino interactions can break the existing correlation between 
$H_0$ and $\sigma_8$. They can increase the value of $H_0$ and simultaneously decrease the value of $\sigma_8$, thus potentially reducing the current tensions between the Planck data and other measurements. However our main conclusions are three-fold.

\begin{itemize}
\item The high multipole polarisation data prefer LCDM-like models, though they do also predict a smaller value for $\sigma_8$ than LCDM. 
\item The DM-$\nu$ interactions do help to reduce the tension between the CMB and weak lensing estimates~\cite{Heymans:2012gg, Erben:2012zw, Hildebrandt:2016iqg} of the $\sigma_8$ value, whatever the CMB dataset under consideration. This is particularly true when $\neff$ and/or $\mnu$ are kept as free parameters. However $\neff$ remains compatible with the Standard Model value, unless one also adds the Cepheids measurements. 
\item DM-$\nu$ interactions can also help to reduce the tensions between the CMB and Cepheid measurements of the Hubble constant, if one disregards the high multipole polarisation data. The combination of the CMB $+$ Cepheid datasets leads to a Hubble rate value of about $72.1^{+1.5}_{-1.7} \ \rm{km} \, \rm{s}^{-1} \, \rm{Mpc}^{-1}$ when  $\neff$ is free to vary (and $71.9^{+1.6}_{-1.8} \  \rm{km} \, \rm{s}^{-1} \, \rm{Mpc}^{-1}$ when both $\neff$ and $\mnu$ are free). 
Under these conditions, $\neff$ can become as large as $\neff=3.54 \, {\pm 0.20}$ or $\neff=3.56\,_{-0.26}^{+0.19}$ if $\mnu$ can vary. In the latter case, we find that the sum of neutrino masses could reach up to 0.87 eV but the $\sigma_8$ parameter  remains too high to reduce both the $H_0$ and $\sigma_8$ tensions simultaneously.  
\end{itemize}

Finally we note that whatever the datasets used and hypothesis that we made, the DM-$\nu$ elastic scattering cross section cannot exceed $\sigma_{\rm{DM}} \lesssim 3 \ 10^{-31} - 6 \ 10^{-31} \ (m_{\rm{DM}/\rm{GeV}}) \ \rm{cm^2}$. 

To conclude, DM-$\nu$ interactions do not enable to solve both the $H_0$ and $\sigma_8$ tensions simultaneously, but they can reduce them slightly nonetheless, if we ignore the high multipole polarisation data. Furthermore the combination of the low multipole and Cepheid data \cite{R16} show that such interactions have the potential to solve the $H_0$ tension, if we ignore the $\sigma_8$ tension. Should there be a good reason to ignore the high multipole polarisation data, one could potentially establish a link between the DM abundance and the neutrino masses \cite{Boehm:2006mi,Ma:2006km,Farzan:2009ji,Farzan:2010mr,Arhrib:2015dez}. 
The DESI~\cite{Levi:2013gra} and Euclid\footnote{See http://sci.esa.int/euclid/.} surveys should be able to determine whether such relatively large interactions were present in the early Universe \cite{Escudero:2015yka}. Such high values of the $u$ ratio would question our understanding of structure formation, as it is expected that there would be little satellite companions left in the Milky Way \cite{Boehm:2014vja}.

\section*{Acknowledgement}
The authors would like to thank J.L. Bernal, M. Escudero, M. Gerbino, A. Riess, L. Verde and R. Wilkinson for very useful discussions.   
EDV acknowledges support from the European Research Council in the form of a Consolidator Grant with number 681431. This research was supported in part by Perimeter Institute for Theoretical Physics. Research at Perimeter Institute is supported by the Government of Canada through Industry Canada and by the Province of Ontario through the Ministry of Economic Development and Innovation.

\end{document}